\documentclass[preprint,11pt]{aastex}

\newcommand{\eg}{e.g.,\ }
\newcommand{\etal}{et~al.\ }

\begin{document}

\title{Confirmation of SBS 1150+599A As An Extremely Metal-Poor 
Planetary Nebula}

\author{George H. Jacoby\altaffilmark{1}}
\affil{WIYN Observatory, P.O. Box 26732, Tucson, AZ, 85726}
\email{gjacoby@wiyn.org}

\author{John Feldmeier}
\affil{Case Western Reserve University, Dept of Astronomy, 10900 Euclid Ave,}
\affil{Cleveland, OH 41406}
\email{johnf@eor.astr.cwru.edu}

\author{Charles F. Claver\altaffilmark{2}}
\affil{Kitt Peak National Observatory, National Optical Astronomy Observatory}
\affil{P. O. Box 26732, Tucson, AZ, 85726}
\email{cclaver@noao.edu}

\author{Peter M. Garnavich}
\affil{Physics Department, University of Notre Dame, Notre Dame, IN 46556}
\email{pgarnavi@miranda.phys.nd.edu}

\author{Alberto Noriega-Crespo}
\affil{SIRTF Science Center, Caltech 220-6, Pasadena, CA, 91125}
\email{alberto@ipac.caltech.edu}

\author{Howard E. Bond}
\affil{Space Telescope Science Institute, 3700 San Martin Dr, Baltimore, 
MD 21218}
\email{bond@stsci.edu}

\author{Jason Quinn}
\affil{Physics Department, University of Notre Dame, Notre Dame, IN 46556}
\email{jquinn@nd.edu}

\altaffiltext{1}{The WIYN Observatory is a joint facility of the
University of Wisconsin-Madison, Indiana University, Yale University,
and the National Optical Astronomy Observatory.}
\altaffiltext{2}{Kitt Peak National Observatory, National Optical
Astronomy Observatory, which is operated by the Association of
Universities for Research in Astronomy, Inc. (AURA) under cooperative
agreement with the National Science Foundation.}

\begin{abstract}
SBS 1150+599A is a blue stellar object at high galactic latitude 
discovered in the Second Byurakan Survey.
New high-resolution images of SBS 1150+599A are presented, demonstrating
that it is very likely to be an old planetary nebula in the galactic halo,
as suggested by Tovmassian et al (2001). An H$\alpha$ image taken with
the WIYN 3.5-m telescope and its ``tip/tilt" module reveals the
diameter of the nebula to be $9\farcs2$, comparable to that estimated
from spectra by Tovmassian \etal Lower limits to the central star
temperature were derived using the Zanstra hydrogen and helium methods
to determine that the star's effective temperature must be $>$68,000K
and that the nebula is optically thin. New spectra from the MMT and FLWO
telescopes are presented, revealing the presence of strong [\ion{Ne}{5}]
$\lambda3425$, indicating that the central star temperature must be
$>$100,000K. With the revised diameter, new central star temperature,
and an improved central star luminosity, we can constrain photoionization
models for the nebula significantly better than before. Because the
emission-line data set is sparse, the models are still not conclusive.
Nevertheless, we confirm that this nebula is an extremely metal-poor
planetary nebula, having a value for O/H that is less than 1/100
solar, and possibly as low as 1/500 solar.

\end{abstract}


\section{Introduction}
Planetary Nebulae (PNe) in our galactic halo provide a unique insight into
the mechanisms of stellar evolution at low metallicities and old ages.
Since many chemical abundances of the progenitor star are not changed
during stellar evolution, halo PNe may probe the metallicity of the
Galactic halo at the formation epoch of their progenitors (Torres-Peimbert
\& Peimbert 1979).  Because of their rarity, however, only a few halo
or Type IV PNe have been discovered (e.g., Howard, Henry, \& McCartney
1997), where in contrast, there are over a thousand disk PNe known (e.g.,
Acker et al. 1992).

In this paper, we report our study of a likely new halo PN:
SBS 1150+599A (PN G135.9+55.9).  Garnavich \& Stanek (1999)
obtained a spectrum of SBS 1150+599A (11:53:24.73 +59:39:57.0 J2000)
along with 2 other cataclysmic variable (CV) candidates identified in
the stellar object list of the Second Byurakan Sky Survey (SBS; Balayan
1997). Of these 3 stars, SBS 1150+599A was uncharacteristic, having
narrow emission lines from the Balmer series as well as \ion{He}{2}
$\lambda4686$ and [\ion{Ne}{5}] $\lambda3425$. Because the object was
small and unresolved, Garnavich \& Stanek suspected it to be a young PN
or symbiotic star. Subsequently, Tovmassian et al. (2001; TSCZGP) obtained
a series of spectra of this object to evaluate its nature. They confirmed
some of the emission characteristics briefly noted by Garnavich \& Stanek
(1999) and they attempted to derive a number of astrophysical quantities
by modeling the photoionization properties of the nebula. Remarkably,
TSCZGP found that the nebula had to be extremely oxygen-poor, despite
the relatively few constraints on their models. In fact, the nebula would
be the most oxygen-poor PN known by about a factor of 10, implying that
it derived from one of the most oxygen-poor stars.

Because this object is so extreme, we wished to verify that it was, in
fact, a genuine PN. If it was another class of object (\eg nova shell,
galaxy, cataclysmic variable), then the TSCZGP abundance models would
be suspect.  If the object was found to be a PN, then we hoped to improve
the abundance models by adding additional observational constraints.
In Section 2, we describe the high spatial resolution images we obtained
with the VATT and WIYN telescopes that confirm the PN hypothesis, as
well as additional spectra obtained at the MMT. In Section 3, we
derive the properties of the PN. In Sections 4 and 5, we discuss our
photoionization model and abundance analysis results.

\section{Observations}
\subsection{Imaging}

If SBS 1150+599A has an extended nebula, it must be small on the sky.
It must have appeared stellar on the SBS survey, or the nebular nature
would have eliminated it from consideration as a CV.  SBS 1150+599A
appears stellar on the POSS-II images available in the Digital Sky
Survey\footnote{The Digitized Sky Surveys were produced at the Space
Telescope Science Institute under U.S. Government grant NAG W-2166.},
and was classified as unresolved in the SBS (Bicay \etal 2000).  (Note
that there is a well-resolved galaxy about 2' north referred to as SBS
1150+599B.) Based on the extent of the spectral lines along the slit of
their spectra, TSCZGP estimated the diameter of the nebula to be $10''$.
Unless the seeing was very good and the spectrograph imaged very well,
this extent was likely to be an overestimate. A direct image taken in
good seeing conditions would clarify the nature of the object further,
and would provide a direct measurement of the nebula diameter.

The nebula diameter serves us in two ways. First, it is an important 
parameter in estimating the distance to the PN.  The Shklovsky method 
(Shklovsky 1956), while notoriously inaccurate, is still a 
competitive tool for measuring distances to random PNe, especially 
when the observational data are few.  Second, the diameter provides 
an upper limit to the radius of integration for the photoionization 
models, essentially constraining the mass of the PN.

We first imaged SBS 1150+599A at the Vatican Advanced Technology Telescope
(VATT) in March 2001. The H$\alpha$ image showed an extended object, but with
a seeing disk of $1\farcs2$ (FWHM), the object diameter could not be measured
accurately. A continuum image taken at 6670 \AA\  demonstrated that a
stellar component existed that was surrounded by an H$\alpha$ nebulosity.
Figure 1 illustrates the VATT images, after being deconvolved using the 
Lucy-Richardson algorithm.

More recently, we imaged SBS 1150+599A at the WIYN 3.5-m telescope 
on Kitt Peak on UT 02 April 2002. A series of 4 exposures were 
co-added, totaling 20 minutes of integration through a narrow-band 
filter (FWHM of 71 \AA) centered on H$\alpha$. The filter 
bandpass includes the [\ion{N}{2}] lines at $\lambda\lambda6548,6583$,
but the spectra indicate that any contribution from
nitrogen emission is small enough to be ignored. Thus, the entire
flux from the nebula can be considered H$\alpha$ emission, plus
some contribution from ionized helium (see Section 4).

The WIYN images were obtained with the WIYN Tip-Tilt Module (WTTM) that
compensates for image motion arising from telescope shake, 
guider errors,
and atmospheric effects (Claver et al. 2002).  The central star of SBS
1150+599A was used as the guide star, with corrections for image motion
being applied at a rate of 200 Hz. The resulting images have a FWHM of
0\farcs53, allowing a clean separation of the central star from the nebula,
an accurate determination of the PN diameter, and illustrating that the
PN has definite structure that departs from a simple sphere or shell.
Figure 2 shows the WIYN image of the PN, both the original and
with the central star subtracted away using several of the
field stars as a guide to the point-spread function (see Section 3).
Figure 3 shows the same image as Figure 2 (right), but as a contour map
to illustrate the morphology of the faint outer parts of the nebula.

With the fine scale ($0\farcs113$/pixel) and good seeing of the WTTM imager,
we are able to measure angular scales quite accurately.  Unfortunately,
there is no evident edge to the nebula; it simply fades into the sky
background at very low surface brightness. Consequently, the diameter is
not easily defined. One estimator of the diameter 
was proposed by Dopita et al. (1996) as the diameter enclosing 85\% of 
the total flux of the nebula. For SBS 1150+599A, this diameter, 
$D(85)$, is $5\farcs0$. Clearly, though, the nebula extends well beyond this 
diameter, and by integrating to $\sim$99\% of the total enclosed flux, 
we derive a diameter, $D(total)$, of $\sim9\farcs2$.  While it is unclear 
what definition of diameter TSCZGP adopted, their value of $10''$ 
is comparable to our latter estimate.

Morphologically, SBS 1150+599A is neither a classic shell nor a uniformly
filled sphere, but appears almost square (see contour map in Figure 3).
If one plots the radial distribution of the emission (Figure 4),
the linear fall-off indicates a slowly declining density with radius.
In addition, there exists an enhancement that appears somewhat bipolar,
perhaps due to a string of ``fast low-ionization emission regions'',
or FLIERs (Balick \etal 1993).  The FLIER interpretation, though,
is untenable since SBS 1150+599A has no low ionization species.
Superficially, SBS 1150+599A looks like a faint version of the PN J320
(Balick 1987).

\subsection{Spectroscopy}

A spectrum of SBS 1150+599A was obtained at the MMT on UT 26 January 1998
and a second spectrum was obtained on 26 March 2001. The latter spectrum
is shown in Figure 5. Both spectra were obtained using a 1\arcsec\ 
slit rotated to the parallactic angle. The exposure times were
900 seconds, and the spectral resolution was $5.5 - 6.0$ \AA\ over
most of the wavelength coverage. The spectrum is remarkable in that the
principal lines are the recombination lines of the hydrogen and helium.
This spectrum, however, extends well into the blue and adds new
information beyond that described by TSCZGP -- the line of
[\ion{Ne}{5}] $\lambda$3426 is extremely strong. It is easily the strongest
collisionally excited line, with a corroborating detection of [\ion{Ne}{5}]
$\lambda$3346. The more familiar lines of [\ion{O}{3}] $\lambda$5007 and
[\ion{Ne}{3}] $\lambda$3868 are extremely weak (see Figure 6).

SBS 1150+599A was also observed on UT 30 May 2000 through
a 3\arcsec\  slit at the Fred L. Whipple Observatory 1.5-m
Tillinghast telescope. This last spectrum
is not very deep, but
it has greater spectrophotometric accuracy than the 
MMT spectra. In particular, the MMT data suffer from second order 
contamination in the red that compromises the flux levels there, leading 
to an underestimate of the H$\alpha$ flux and an unphysically 
small Balmer decrement. The ratio of H$\alpha$/H$\beta$ from the 
FLWO spectrum is 2.86, which we adopt as the observed value 
before de-reddening. This line ratio is higher than reported by TSCZGP, 
who found 2.52 from their spectra.  The remaining line ratios
relative to H$\beta$ are taken from the deeper MMT data.  Since these
lines are much bluer than H$\alpha$, they do not suffer from second
order effects.

Table 1 lists the emission lines seen in the spectra, their heliocentric
radial velocities, and their fluxes relative to H$\beta$.  For the fluxes
we list both the observed values (average of the two MMT spectra, except
for H$\alpha$, which is taken from the FLWO spectrum), and the values from
our model (described in Section 4).  The uncertainties in the fluxes of
the strong lines are typically 3-5\%, and are dominated by the standard
star calibrations and the reduction processes. For the faint lines,
the uncertainties are dominated by the placement of the
continuum. We estimated this uncertainty by measuring the flux assuming
a range of plausible continuum values and determining the average and
variance in the resulting line flux.

The observed Balmer decrement is shallow,
suggesting that the interstellar extinction is small. The observed
fluxes listed in Table 1 have been corrected for an adopted $E(B-V) =
0.03$, based on the maps of Schlegel, Finkbeiner, \& Davis (1998).

We measure the heliocentric radial velocity of SBS 1150+599A 
to be $-194\pm3$ km-s$^{-1}$, in good agreement with TSCZGP who 
report a velocity of $-190$ km-s$^{-1}$.  Thus, we confirm that 
the velocity of this PN is consistent with it being part of 
the Galactic halo population.

\section{Nebula and Central Star Parameters}

\subsection{Central Star Temperature}

We must have a reasonably accurate estimate of the central star
temperature in order to calculate an oxygen abundance. The temperature
becomes a critical parameter because it determines the ionization
state for this low density, optically thin, nebula. Without knowing the
ionization state, a weak [\ion{O}{3}] $\lambda$5007 flux can be evidence for
either a low oxygen abundance, a nebula ionization state in which 
the oxygen is in states much higher (or lower) than doubly ionized, or some 
combination of these.

The temperature of the central star of SBS 1150+599A must be high
for a variety of reasons. Most importantly, the mere presence of
[\ion{Ne}{5}] $\lambda$3426 requires a star of 100,000K or
hotter. Similarly, the relative strength of \ion{He}{2} $\lambda4686$
to \ion{He}{1} $\lambda5876$ requires a central star temperature of at
least 80,000K. The continuum visible in our own spectra, and that of
TSCZGP is also consistent with a hot central star. In addition,
we have identified
a soft x-ray source in the ROSAT all-sky survey bright source catalog 
RASS-BSC (Voges \etal 1999),
1RXS J115327.2+593959, that has a hardness ratio of $-1.0\pm0.3$,
and is $17\farcs46\pm17$ distant from SBS 1150+599A. If it proves to be
associated with the PN central star, it provides further support for 
a temperature of 100,000K or hotter (Guerrero, Chu, \& Gruendl 2000).

We can also derive a lower limit for the stellar temperature 
using the Zanstra method which depends only on the ratio between the 
stellar flux and the nebular flux (see,
for example, Pottasch 1984).  The usual Zanstra relationship between
central star $V$ magnitude and H$\beta$ flux can be restated for any
central star bandpass and any Balmer emission line. We adopt a ``red"
instrumental magnitude based on the star's H$\alpha$ flux that we
relate to a $V$ magnitude assuming a hot black body for the star. We
also measure the nebula's H$\alpha$ instrumental flux directly and use
the observed de-reddened Balmer decrement of 2.81 to transform 
to H$\beta$.

The first step is to measure the central star magnitude, a process that
is complicated by the presence of the nebula, but simplified if we first
subtract the image of the central star.  To do this, we used the IRAF
version of DAOPHOT (Stetson 1987; Stetson 1992).  We created an
empirical point-spread-function (PSF) from three bright, uncrowded,
stars on the frame.  We then fit the inner core of the nebula to the
PSF, using the NSTAR task within DAOPHOT.  The central star subtracts
away well, with minimal residuals, as can be clearly seen in Figure~2.

When a PN is optically thin, ionizing radiation escapes and the derived
Zanstra temperature is an underestimate.  If one can derive the Zanstra
temperature from emission-lines having a range of ionization limits,
such as H, He, and He$^+$, one can directly test whether the nebula is
optically thin or not.  In the thin case, the derived Zanstra temperature
increases with ionization levels. For SBS 1150+599A, we can use the
\ion{He}{2} $\lambda4686$ and H$\beta$ lines to compare our 
hydrogen Zanstra temperature to that for helium. As noted by 
Kaler \& Jacoby (1989), these two temperatures will be equal 
in the optically thick case.  For this nebula, though, 
the hydrogen temperature of 25,500 K is
unrealistically low.  The HeII temperature of 68,600 K is reasonable for
a PN central star, but it, too, is very likely to be an underestimate
because of the strong emission of [\ion{Ne}{5}].  Thus, the
central star must be hotter than $\sim$68,000 K.

The photoionization models described in Section 4 blend all the above
information into a more quantitative formulation. We will show that a
reasonable assessment of the central star temperature is $\sim100,000$K.

\subsection{Central Star Luminosity}

For a very low density nebula like this one, which is fully
ionized throughout, the central star luminosity is not a critical
parameter. Nevertheless, it does have secondary impact on the ionization
structure of the nebula and helps to constrain the photoionization models
if it can be measured.

In our WIYN image, the central star contributes 14.3\% of the total
flux in the nebula plus star. We adopt the nebular flux from TSCZGP
(see Section 3.3). Thus, the stellar flux, integrated over the 71 \AA\
bandpass of the filter, is $5.78 \times 10^{-15}$ ergs cm$^{-2}$ s$^{-1}$,
which is equivalent to $m_{6563} \sim 18.6$. For a hot black body,
this corresponds to  $m_{\mbox{v}}\sim 18.3$.  A direct measure from
the MMT spectrum yields $m_{\mbox{v}} =18.2$. We will use the average
magnitude of 18.25, combined with the distance (see Section 3.4) to derive
the absolute $V$ magnitude for the star.  Correcting for extinction,
we adopt  $m_{\mbox{v$_{0}$}} = 18.16$.

\subsection{Nebular Flux}

TSCZGP report the H$\alpha$ nebular flux through a 3\arcsec\  slit
(presumably centered on the central star) of $3.01 \times 10^{-14}$ ergs
cm$^{-2}$ s$^{-1}$. After removing the contribution to our image from
the central star, we find that a 3\arcsec\  slit samples 74.5\% of the
total integrated nebular flux. Thus, the total nebular emission in the
H$\alpha$ bandpass is $4.04 \times 10^{-14}$ ergs cm$^{-2}$ s$^{-1}$.
Photoionization models indicate that the contribution from
HeII to the H$\alpha$ and H$\beta$ fluxes is about 5\%. Thus, the total
H$\alpha$ flux is $3.83 \times 10^{-14}$ ergs cm$^{-2}$ s$^{-1}$ and 
after correcting for extinction, the H$\alpha$ and H$\beta$ fluxes are
$4.1 \times 10^{-14}$ and $1.47 \times 10^{-14}$ ergs cm$^{-2}$ s$^{-1}$,
respectively.

\subsection{Distance}

We wish to have a reasonable estimate of the distance to help construct
photoionization models that match sensible values for nebular
and stellar fluxes, and nebular and stellar masses.  Unfortunately,
distances for Galactic PNe are rarely accurate. For SBS 1150+599A, we
face the usual uncertainties in this field (\eg Kwok 2000), and must 
resort to statistical distance techniques.

We start with parameters chosen for the Shklovsky method by TSCZGP
in which they derive a distance of 20.4 kpc. We revisit several of these
parameters below. That is, we adopt the nebular mass to be 0.2 M$_\sun$,
the filling factor to be 0.2, and the nebular electron temperature to
be 12,000 K. Now, if we use our measured diameter, $D(total)$ of $9\farcs2$,
and the H$\beta$ flux listed above, we derive a distance to 
SBS 1150+599A of 23.6 kpc. 

We now apply several corrections.  The nebular temperature is an important
factor. Without metals to cool the nebula, the electron temperature could
be quite high, and may approach 30,000 K. We retain the filling factor
at 0.2; although there is no evidence that the nebula is anything but
fully filled, the density distribution cannot be uniform (see Figure 4).
The filling factor is always uncertain to some degree, but it enters into
the distance calculation only very weakly. The nebular mass is probably
the most important unknown. If the progenitor star is truly
very metal-poor and represents an epoch early in the formation history
of the galaxy, the star must have had a main sequence mass less than
the Sun. In that case, the nebula is also likely to be very low mass.
The most comparable object in the Galaxy is probably the PN
in the globular cluster M15 (Ps 1) which has a nebular mass of
$\sim 0.05$ M$_\sun$ (Buell et al 1997). But, adopting the mass of Ps
1 has no other independent justification.

With these corrections, and assuming a nebular mass between 0.05 and
0.3 M$_\sun$, we derive a range of Shklovsky distances of 11.5 to 23.5 kpc. 
The biggest factor in broadening the range of derived distances is the mass
of ionized gas, and thus far, we can only assume a value for this parameter.
At the smallest distance, though, the central star luminosity drops to 
the point where its mass becomes
$\sim$0.54 M$_\sun$, a value considered below the threshold for 
producing a visible PN (Jacoby \etal 1997; Alves, Bond, \& Livio 2000).

An alternative and new method for deriving distances to PNe is proposed
here based on the surface brightnesses of PNe.  In their study of LMC PNe,
which are all at the same distance, Stanghellini et al (2002) noted that
there exists an excellent correlation between the nebula's H$\alpha$
surface brightness and its linear radius. Because surface brightness is
independent of distance, but linear radius depends directly on distance, a
correlation between these two parameters can be used to infer the distance
from simple measurements of the surface brightness and angular radius.
The correlation has a $1\sigma$ dispersion of $\sim16$\% in linear
radius, and therefore, if it holds in general, should yield distances
that are quite accurate.

For elliptical and round PN, such as SBS 1150+599A, the surface brightness
-- radius (SBR) relation follows the form:

$${\rm Log} (R_{phot,pc}) = -4.908 - 0.324\  {\rm Log} (SB),$$

\noindent
where $R_{phot,pc}$ is the linear ``photometric'' radius of the
nebula in parsecs, and $SB$ is the H$\alpha$ surface brightness
in ergs-cm$^{-2}$-s$^{-1}$-arcsec$^{-2}$.  The ``photometric''
radius corresponds to half of $D(85)$, the diameter enclosing 85\%
of the total flux of the nebula, and $SB$ is the total flux divided
by $\pi R_{phot,\theta}^2$. $R_{phot,\theta}$ is the angular
``photometric'' radius in arcsec, which relates in the obvious way
to $R_{phot,pc}$ such that the distance, $d(pc)$, is $206265 R_{phot,pc} /
R_{phot,\theta}$. We note that $R_{phot,\theta}$ is defined to be
the diameter as measured in [\ion{O}{3}], but because of the extreme nature
of this nebula, we have substituted the diameter as measured 
at H$\alpha$.  This could introduce a systematic error in the 
distance derivation.

For SBS 1150+599A, Log$(SB) = -14.67$ and $R_{phot,\theta} = 2.5$.
The distance would follow directly, but we must make one adjustment to
the correlation identified by Stanghellini et al (2002)
before proceeding. Because the LMC PNe used to derive the correlation
have relatively normal chemical compositions, their nebular electron
temperatures are generally $\sim 12,000$K, whereas SBS 1150+599A is
so metal-poor that its electron temperature may be as high as $\sim
30,000$K. The Balmer emission drops with increasing temperature by
about 0.25 dex over this range. Consequently, we must raise the measured
value of Log$(SB)$ to $-14.41$ to place this unusual object onto the
relationship. Then, we derive that $R_{phot,pc} = 0.58$ pc.  This radius
corresponds to a distance of 47.6 kpc.  However, this distance is not a
comfortable one either, as now the nebular mass must be $\sim$2 M$_\sun$
to meet the nebular luminosity demand, an unlikely value for an old halo
object. It seems likely that SBS 1150+599A is sufficiently bizarre that
it does not obey the SBR relation.  Note that the SBR 
relation, though, is only defined for PNe having surface brightnesses 
a factor of 10 or more larger, and extrapolating below the defined region 
is not justified.

Consequently, we do not have a reliable distance. Without
any other justification than consistency, we simply adopt the
Shklovsky distance as modified above, but assuming the commonly
used nebular mass of 0.2 M$_\sun$. The resulting distance then, is
20.0 kpc. The Galactic latitude of SBS 1150+599A is 56 degrees, and so
it is $\sim16.6$ kpc above the Galactic plane. Regardless of the
uncertainty in the distance, SBS 1150+599A must be well out of the 
plane of Milky Way, and is clearly a halo object.

At this adopted distance, the central star has $M_V = +1.66$. The
central star magnitude serves as an additional observational constraint
to be matched by the photoionization model. For a hot ($\sim100,000$K)
black body, this magnitude corresponds to a luminosity of $\sim$5000
L$_\sun$.  In addition, the nebular $H\alpha$ luminosity at this
distance provides another model constraint and is $1.8 \times 10^{33}$
ergs-cm$^{-2}$-s$^{-1}$.

\section{Photoionization Model}

We use the Cloudy photoionization modeling code, as recently released
(Version 96 Beta 3; Ferland 2001). We assume the nebula has spherically
symmetric geometry and that the central star can be represented as a
black body.  We have not considered the case of non-spherical
geometry because the spherical models already reproduce the constraints
adequately, and the true three-dimensional geometry is an unknown that
cannot be constrained by the observations.

In modeling the nebula, we start by defining
the ionization structure of the nebula that is implied by the flux ratios of
the three neon lines.  These lines provide the only insight to the ionization
structure; no other elemental emission lines are seen arising from more than
one ionization stage.  Consequently, the weak line of
[\ion{Ne}{3}] $\lambda$ 3868 assumes a very important role.  Of the unknown
parameters in the model, the ionization structure depends primarily on
(1) the central star temperature, (2) the central star luminosity, and
(3) the nebular density.  We adopt an electron density distribution
that yields a linearly falling surface brightness as a function of
radius, as suggested by Figure 4, and scale this distribution such
that the nebular luminosity is achieved.  The absolute magnitude of
the central star further constrains these parameters.  Unfortunately,
the uncertain distance to the nebula enters into these considerations
because it affects both the nebular and stellar luminosities.

The remaining unknowns in the model are primarily the abundances, and
these are constrained by the observed emission-line ratios relative to
H$\beta$.  Fortunately, the abundances have little impact on the 
nebular luminosity or the ratio of the [\ion{Ne}{5}] to [\ion{Ne}{3}] lines.
This is an important point because
we can only derive abundances for the 3 elements producing emission lines:
helium, neon, and oxygen.  For the other important elements (sulfur,
argon, nitrogen, and carbon), we can only derive upper limits beyond
which an emission-line would have been seen in our spectra.

With the few lines listed in Table 1, it is not difficult to develop 
a model that reproduces the constraints and line ratios perfectly.
In fact, there are many solutions and our first success
yielded a value for O/H that was $300\times$ below the Solar value, and a
value for Ne/H that is  $10\times$ below Solar.  That model, though, was not
adjusted to meet the upper limit requirements on elements with unseen
emission. Instead, we simply set the abundances of the unseen elements
to be consistent with those for typical Galactic PN, but reduced
in proportion to the Ne/H measured for SBS 1150+599A. 
Consequently, the gas electron temperature of this model was very
high at 23,700K. At this temperature, very little oxygen is needed to
produce the observed strength of [\ion{O}{3}] $\lambda5007$.  We could also
have adopted an abundance set that is in accord with O/H instead of Ne/H,
and the electron temperature would have exceeded 30,000K and O/H would
have been more than $500\times$ smaller than Solar.

Our final model, which is probably the most conservative, is summarized
in Table 2. It is based on the maximum values for S/H, Ar/H, N/H, and
C/H that can be accommodated before emission from these elements would
be seen in our spectra.  Because this model contains the maximum number
of coolants, the electron temperature is reduced to 17,600K. At this
lower temperature, the value for O/H must be increased to $105\times$ 
smaller than Solar in order to match the strength of [\ion{O}{3}] 
$\lambda5007$.

We caution the reader on two crucial points. First, photoionization
models based on the few lines that we can observe in SBS 1150+599A are
not unique.  Our derived abundances for oxygen and neon can vary by
factors of a few while matching the observed constraints perfectly.
Secondly, we reproduced the models to meet the observational data,
which itself, is uncertain. As evidence of this point, note the values in
Table 1 for the fainter Balmer lines. The observed values for H$\gamma$,
H$\delta$ and H$\epsilon$ are systematically smaller than the models
or theory predict. The measurements of the data are reliable, but the
data may be systematically too faint towards the blue. 
A systematic error of this kind, though, is in conflict with the good
agreement seen between the observed stellar continuum and that expected
from a hot star. 

Nevertheless, because the ionization structure of our models depends
strongly on the weak line of [\ion{Ne}{3}] $\lambda 3868$, we ran
an experimental model in which the intensity of this critical line
was reduced by half.  The ionization structure can now be matched by
increasing the central star effective temperature to 107,000K. The most
significant changes that follow are (1) the neon abundance is reduced by
45\%, and (2) the oxygen abundance is increased by 31\% to $76\times$ below
the Solar value.  Consequently, the Ne/O ratio drops from 4.3 in our
baseline model to 2.3 in the low-$\lambda3868$ model.  We consider these
deviations as indicative of the degree of uncertainty in the analysis.

As a final note, the models indicate that the Balmer line measurements
are contaminated by their neighboring He$^{++}$ lines by about 5\%.
This effect was taken into account in the luminosity and the line ratio
measurements adopted throughout.

\section{Discussion}

This study was motivated by the apparently very low oxygen abundance
in the nebula as first reported by TSCZGP. If correct, the object is
unique and warrants verification. In fact, we confirm the suspicion
that SBS 1150+599A is an extremely metal-poor PN, having an oxygen
abundance at least 100$\times$ smaller than that of the Sun.  We can
build models that lead to an oxygen abundance as low as 500$\times$
smaller than that of the Sun, as suggested by TSCZGP, but we favor the
less extreme situation. Still, SBS 1150+599A appears likely to have
a lower O/H than any other PN currently known, being a factor of
$5\times$ lower than Ps 1 in the globular cluster M15 (Howard et al 1997).

Is the progenitor of SBS 1150+599A consistent with being an old metal-poor
star?  Assuming that the very low O/H of the nebula provides an indication
of the progenitor star's Fe/H, then the implication is that SBS 1150+599A
did descend from a very low metallicity star. Its chemcial composition
would be more consistent with that of an old globular cluster or halo
star than the usual intermediate age thick disk population with which
PNe are usually associated.

As a further test of the origin of this PN, we derive the mass of the
central star by interpolating into the evolutionary tracks of Bl\"ocker
(1995) using our derived effective temperature and luminosity.
The central star mass of 0.61 M$_{\sun}$, when combined with a
nebular mass of 0.2 M$_{\sun}$, yields an initial system mass of 0.8
M$_{\sun}$. While this value is consistent with an old progenitor,
the age, and the stellar and nebular masses are all dependent on
a very uncertain distance. Furthermore, the central star mass is
far too high to have originated from a low-mass progenitor under any
reasonable initial-to-final mass relationship (Weidemann 2001). Unless
the distance is too long by a factor of 2 or more, the progenitor of
SBS 1150+599A must have been a single star with a zero-age mass of $>$
1.3 M$_{\sun}$. Alternatively, the progenitor could have formed from a
coalesced binary pair (Jacoby \etal 1997; Alves \etal 2000).  Only the
latter possibility is consistent with a very old star originating in the
Galactic halo.

SBS 1150+599A is also likely to be an old, evolved PN.  With a linear
diameter of 0.87 pc, and assuming an expansion velocity of 20 km-s$^{-1}$,
the expansion age of this PN is 21,000 yrs.

Perhaps the most intriguing results are those for the abundances
of neon and oxygen. The value of Ne/O that we find is 4.3. Typical
Galactic PN have values of 0.26, and there is generally a very tight
relationship between Ne/H and O/H (Henry 1989) that this object strongly
violates. While Henry (1989) did not have any objects in his sample
with O/H as low as in SBS 1150+599A, he did note that BoBn~1, also a halo
PN, had an anomalously large value of Ne/O. Howard et al (1997) 
find that Ne/O for BoBn~1 is 0.8.

BoBn~1 also has a high N/O ratio (1.5).  We have only an upper limit for
N/H, and the value is very high, leading to the relation that N/O $<$
52. While very likely to be a valid statement, this inequality carries
little weight. A real measure of a nitrogen line would be extremely
valuable, and this measurement is feasible in the UV.  With these
very high N/H abundances, the UV lines of [\ion{N}{5}] $\lambda1240$ and 
[\ion{N}{4}] $\lambda1486$ are predicted to be $\sim11\times$ 
stronger than H$\beta$.

\section{Conclusions}

We have developed a fairly consistent picture in which SBS 1150+599A is an
old PN derived from a progenitor in the Galactic halo.
The central star must be $\sim100,000$K and the nebula must be very
oxygen-poor.

The picture, though, is not completely satisfying because the observational
data are so limited. Future studies should include UV observations; that
region of the spectrum is likely to pay the most dividends in constraining
the abundances for carbon and nitrogen. These elements control the cooling
rate of the nebula, and thus, the abundance of oxygen cannot be
measured accurately until that rate is determined. Alternatively,
one can attempt to measure the electron temperature directly from the 
line ratio of [\ion{O}{3}] $\lambda4363$ to $\lambda5007$, but doing 
so would be a difficult observation due to the low fluxes of these
lines.

Nevertheless, the evidence is very strong
that SBS 1150+599A is extremely underabundant in oxygen, and possibly
is the PN with the lowest oxygen abundance. There is some chance that
oxygen has somehow been transformed to neon within the progenitor star,
thereby explaining the very high Ne/O and N/O ratios.  That scenario has
been proposed by Howard et al (1997) for the anomalous halo PN BoBn~1. If
so, then the composition of this PN no longer represents the original
composition of its progenitor star, and the abundances measured in  SBS
1150+599A become more valuable for studying stellar physics than for
studying Galactic chemical history.  More observational study of this
unique object is clearly warranted.

\acknowledgements
We wish to thank Pete Challis for taking the 2001 MMT spectrum.

\begin{figure}
\epsscale{0.8}
\plotone{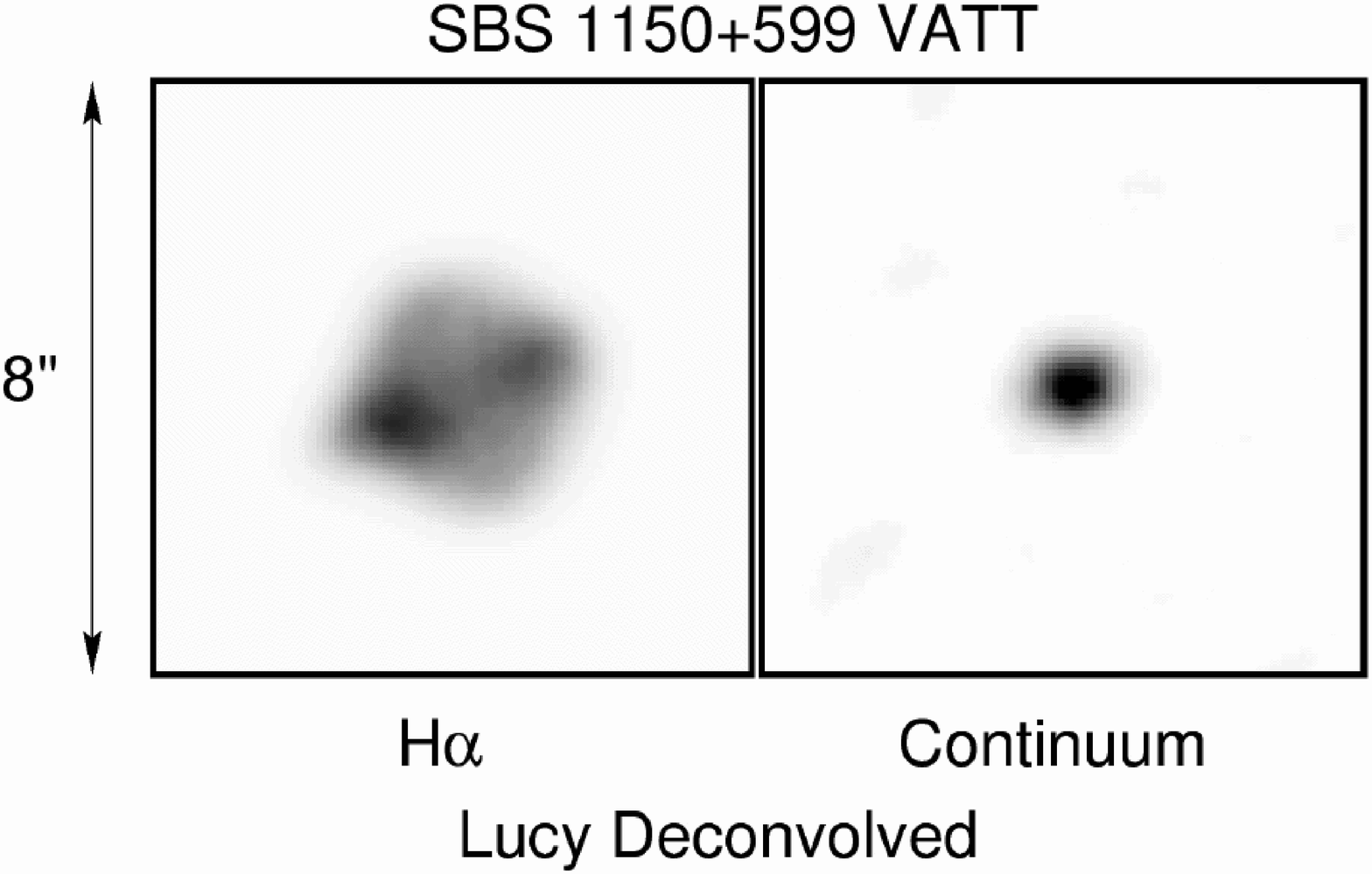}
\caption{VATT images of SBS 1150+599A with the spatial resolution
improved using a Lucy deconvolution algorithm. The left-hand image was taken
with a narrow H$\alpha$ filter while the right-hand image was taken through
a continuum filter centered on 6670 \AA. North is up and east is to the left.}
\end{figure}

\begin{figure}
\epsscale{0.8}
\plotone{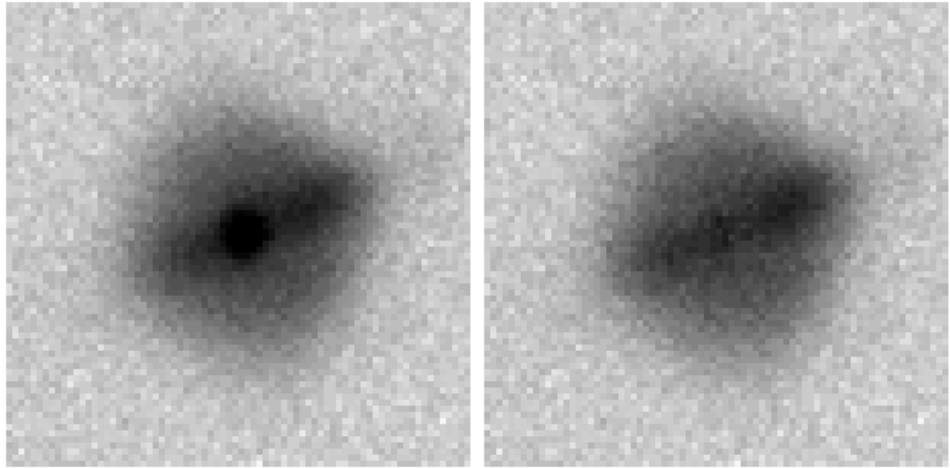}
\caption{The WIYN H$\alpha$ image of SBS 1150+599AA (left) and 
after subtracting
off the central star image (right). The field of view of the image shown
is $\sim 8\farcs5$ on a side. North is up and east is to the left.}
\end{figure}

\begin{figure}
\epsscale{0.8}
\plotone{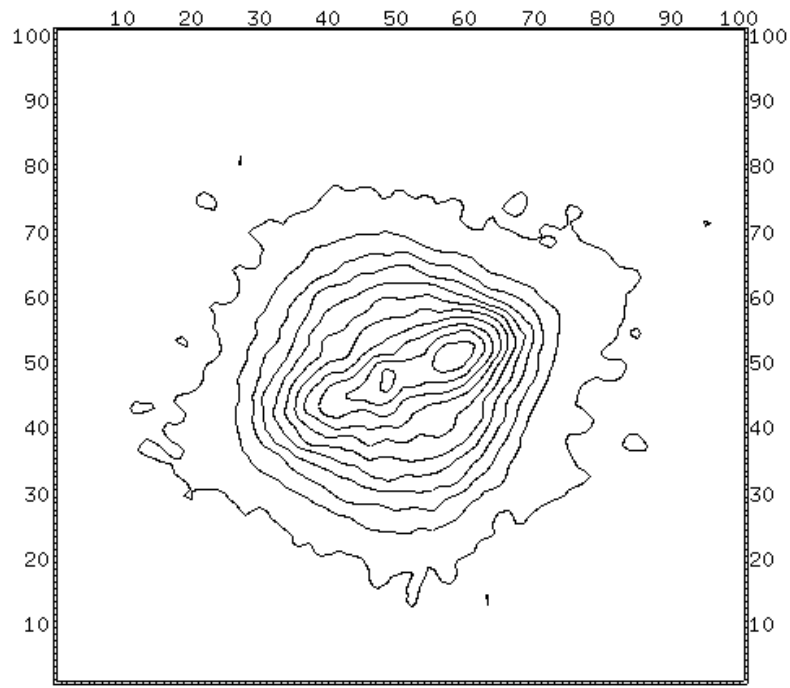}
\caption{A contour map of the nebula, with the central star removed. The
morphology of this nebula shows a curious square outer profile with
a linear enhancement along a diagonal that could be due to a central torus
seen nearly edge-on. The sub-image displayed is 100 pixels ($11\farcs2$) 
on a side.}
\end{figure}

\begin{figure}
\epsscale{0.8}
\plotone{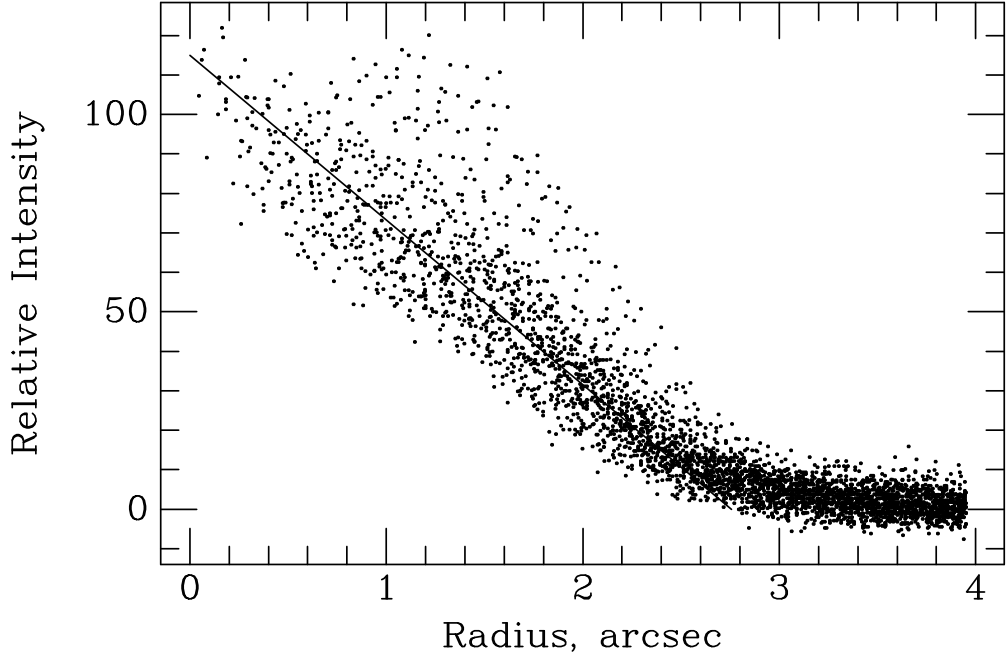}
\caption{The radial flux profile of the nebula shown as points, 
with a linear fit to the relative intensities shown as 
115 - 41.7 R (arcsec).}
\end{figure}

\begin{figure}
\epsscale{0.8}
\plotone{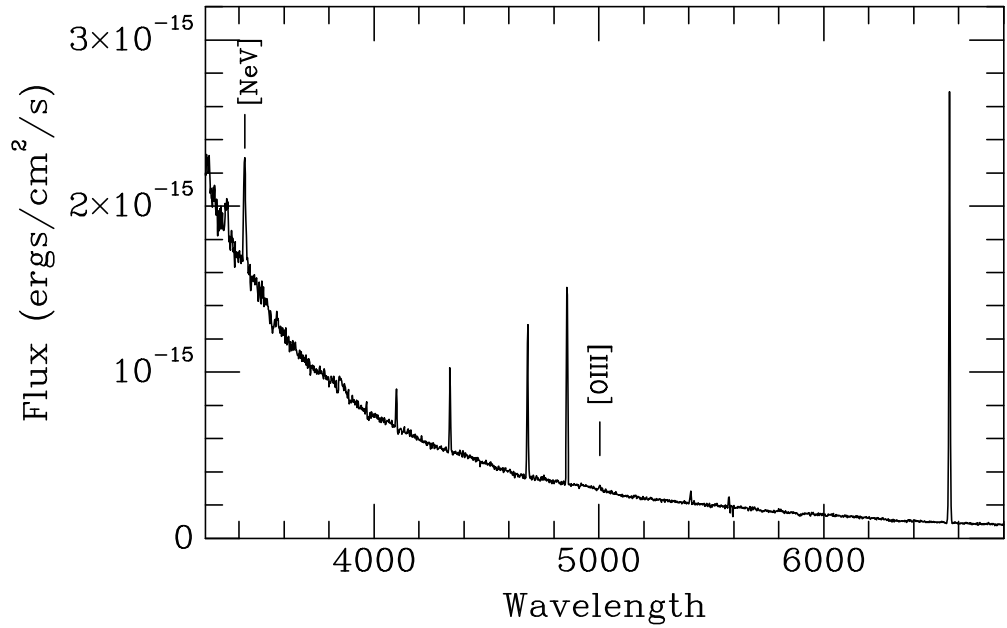}
\caption{The MMT spectrum of SBS 1150+599A showing the near-total lack of
metal lines, with the Balmer lines being dominant. The strong 
[\ion{Ne}{5}] $\lambda$3426 line and the extremely weak  [\ion{O}{3}] 
$\lambda$5007 lines are noted. A strong blue continuum indicates
the presence of a hot central star having $T > 50,000$K.}
\end{figure}

\begin{figure}
\epsscale{0.8}
\plottwo{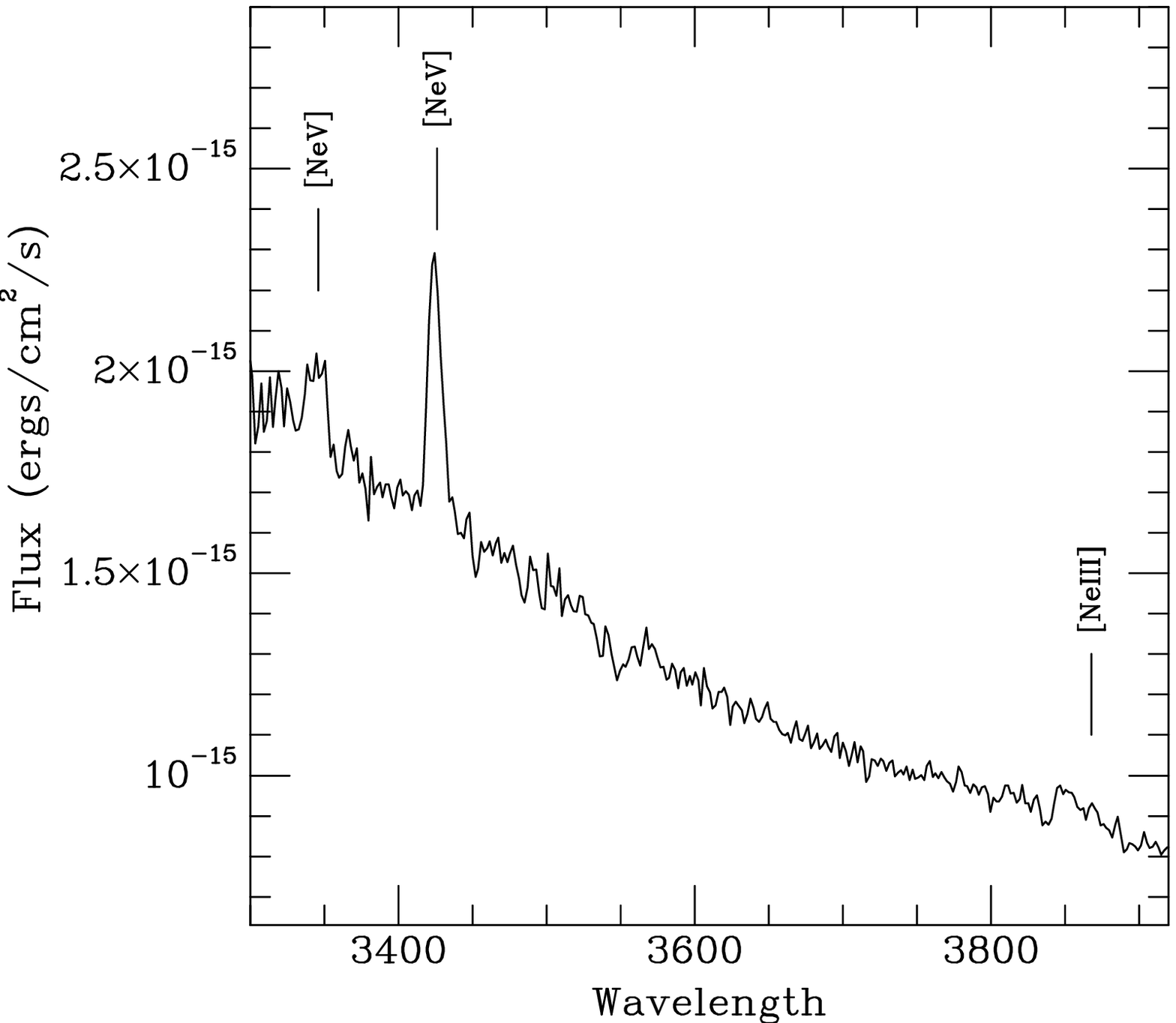}{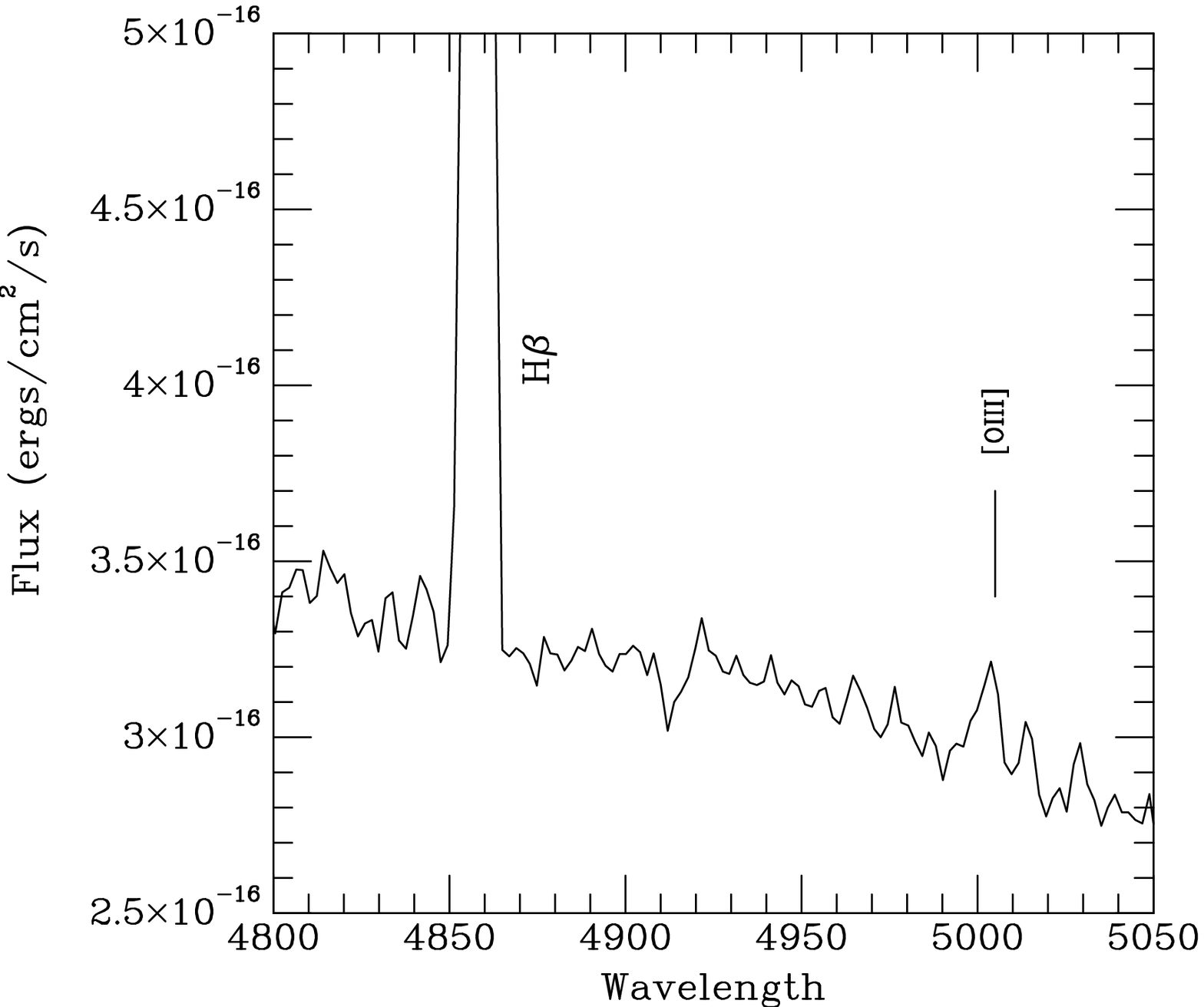}
\caption{Close up views of the regions of the MMT spectrum surrounding
the neon and oxygen lines.  Note the strong detection of [\ion{Ne}{5}]
$\lambda$3426, and the near absence of [\ion{O}{3}] $\lambda$5007}.
\end{figure}

\begin{deluxetable}{lllll}
\footnotesize
\tablecaption{Observed and Predicted Line Fluxes for SBS 1150+599A}
\label{tab:line_fluxes}
\tablewidth{00pt}
\tablehead{
\colhead{Line} & \colhead{ID} & \colhead{Velocity} &
\colhead{Observed Flux$^a$} & \colhead{Model Flux}  \\
\colhead{\AA} & \colhead{ } & \colhead{km-s$^{-1}$} & \colhead{H$\beta$ = 100}
   & \colhead{H$\beta$ = 100} }
\startdata
3345.9  & [\ion{Ne}{5}] 3346       & ...         & $32\pm8$     &  32.0 \\
3424.6  & [\ion{Ne}{5}] 3426       & ...         & $86\pm9$     &  87.5 \\
3860    & [\ion{Ne}{3}] 3868     & ...         & $15\pm7$     &  14.5 \\
3966.2  & H$\epsilon$ 3970 & ...         &  $4\pm1$     &  15.9 \\
4098.78 & H$\delta$ 4102   & $-213\pm23$ & $17\pm2$     &  25.9 \\
4337.43 & H$\gamma$ 4340   & $-208\pm10$ & $39\pm3$     &  47.0 \\
4682.72 & He II 4686       & $-191\pm5$  & $77\pm3$     &  77.2 \\
4858.13 & H$\beta$ 4861    & $-194\pm5$  & $100$        & 100.0 \\
5003.0  & [\ion{O}{3}] 5007      & ...         & $3\pm1$      &   3.1 \\
5407.6  & He II 5412       & $-216\pm20$ & $5\pm1$      &   6.2 \\
6558.67 & H$\alpha$ 6563   & $-193\pm5$  & $281\pm10$   & 276.8 \\
\enddata
\tablenotetext{a}{Line ratios have been corrected for extinction
 value of $E(B-V)=0.03$, and for the $\sim5$\%
 contribution of He$^{++}$ lines to H Balmer lines.}
\end{deluxetable}

\begin{deluxetable}{lcc}
\footnotesize
\tablecaption{Summary of Model Parameters}
\label{tab:model_out}
\tablewidth{00pt}
\tablehead{
\colhead{Parameter} & \colhead{SBS 1150+599A} & \colhead{Solar$^a$} }
\startdata
Central Star Temperature & 101,000 K & \\
Central Star Luminosity  & 5750 L$_{\sun}$ & \\
Central Star Mass        & 0.61 M$_{\sun}$ & \\

Nebular Density$^b$      & 21 cm$^{-3}$  & \\
Nebular Temperature$^c$  & 17,600 K  & \\
Nebular Mass             & 0.2 M$_{\sun}$  & \\

N(He)/N(H)      &   0.066  &  0.098 \\
log(O/H)+12     &   6.93   &  8.93 \\
log(Ne/H)+12    &   7.47   &  8.09 \\
log(N/H)+12     &  $<$8.65   &  8.00 \\
log(S/H)+12     &  $<$7.00   &  7.21 \\
log(C/H)+12     &  $<$8.30   &  8.60 \\
log(Ar/H)+12    &  $<$5.00   &  6.56 \\
Ne/O		&   4.3	     &  0.15 \\
N/O		&   $<$52    &  0.12 \\
\enddata
\tablenotetext{a}{Solar abundances are taken from Clegg's 1992 summary
  of Grevesse \& Anders 1989 and Grevesse et al. 1990, 1991}
\tablenotetext{b}{Electron density averaged over volume}
\tablenotetext{c}{Electron temperature, density weighted and averaged over
volume}
\end{deluxetable}

\end{document}